\documentclass[11pt,twoside]{article}

\usepackage{asp2006}

\begin{document}
\title{Star-Forming Galaxies at \boldmath $z \sim 2$: \\
An Emerging Picture of Galaxy Dynamics and Assembly}   
\author{Kristen L. Shapiro\altaffilmark{1}, Reinhard Genzel\altaffilmark{1,2}, Nicolas Bouch\'e\altaffilmark{2}, Peter Buschkamp\altaffilmark{2}, Giovanni Cresci\altaffilmark{2}, Ric Davies\altaffilmark{2}, Frank Eisenhauer\altaffilmark{2}, Natascha F\"orster Schreiber\altaffilmark{2}, Shy Genel\altaffilmark{2}, Erin Hicks\altaffilmark{2}, Dieter Lutz\altaffilmark{2}, Linda Tacconi\altaffilmark{2}}
   
\altaffiltext{1}{Department of Astronomy, University of California at Berkeley, Berkeley, CA 94720, USA}
\altaffiltext{2}{Max-Planck-Institut f\"ur extraterrestrische Physik, Giessenbachstr.1, D-85748 Garching, Germany}

\begin{abstract} 
In these proceedings, we summarize recent results from our ``SINS" {\it VLT}/SINFONI integral-field survey, focusing on the 52 detected UV/optically-selected star-forming galaxies at $z \sim 2$.  Our H$\alpha$ emission-line imaging and kinematic data of these systems illustrates that a substantial fraction ($\geq 1/3$) of these galaxies are large, rotating disks and that these disks are clumpy, thick, and forming stars rapidly.  We compare these systems to local disk scaling relations and find that the backbones of these relations are already in place at $z \sim 2$.  Detailed analysis of the large disks in our sample provides strong evidence that this population cannot result from a merger-dominated formation history and instead must be assembled by the smooth but rapid inflow of gas along filaments.  These systems will then secularly evolve from clump-dominated disks to bulge-dominated disks on short timescales, a phenomenon that is observed in our SINS observations and is consistent with predictions from numerical simulations.  These results provide new and exciting insights into the formation of bulge-dominated galaxies in the local Universe.
\end{abstract}


\section{Introduction: The SINS Survey}
\label{sec:intro}

The $z \sim 2$ Universe is now known to represent a critical epoch in matter assembly; during this era, both the cosmic star formation rate and the luminous quasar space density are at their peaks \citep[e.g.][]{Fan+01_KLS,Cha+05_KLS}.  The assembly of galaxies is correspondingly rapid, with the total stellar mass density in galaxies increasing from $\sim 15$\% to $50-75$\% it current value between $z \sim 3$ and $z \sim 1$ \citep[e.g.][]{Dic+03_KLS,Rud+03_KLS,Rud+06_KLS}.  Observations of the dynamical and baryonic processes driving this evolution are therefore central to our understanding of galaxy formation.

To this end, we have conducted the SINS survey (Spectroscopic Imaging in the Near-infrared with SINFONI).  The near-infrared integral-field capabilities of SINFONI allow us to probe the two dimensional distribution and kinematics of redshifted H$\alpha$ and other key diagnostic emission lines.  Our selection techniques and large sample size (52 detected UV/optically-selected galaxies, of 62 observed) enable us to probe a representative subset of the population of star-forming galaxies at $z \sim 2$, with an observationally-imposed slight bias towards  more massive and more rapidly star-forming galaxies ($\langle M_* \rangle \sim 3 \times 10^{10}\ M_\odot$, $\langle {\rm SFR} \rangle \sim 50\ M_\odot\ {\rm yr}^{-1}$; F\"orster Schreiber et al.~in preparation).  With SINFONI, we study this population with typical spatial resolution of $0.5-0.6$\arcsec\ ($\sim 4-5$ kpc) and spectral resolution of $70$~km~s$^{-1}$.  A subset of this sample has additionally been observed with adaptive optics and resolved at $0.15-0.3$\arcsec\ ($\sim 1-2$ kpc).

The SINS data reveal a diversity in kinematics and morphologies of these systems, as seen in H$\alpha$ emission (\citealt{For+06_KLS,Gen+06_KLS,Bou+07_KLS}; see also related work by \citealt{Law+07_KLS,Wri+08_KLS}).  The population at this redshift includes large rotating disks, compact dispersion-dominated systems, and several interacting or merging galaxies, each accounting for roughly $1/3$ of the total population (F\"orster Schreiber et al.~in preparation).  Of these, the large rotating disks, discovered and probed for the first time with our SINS observations, have revealed much about the nature and evolution of high-redshift galaxies.

In these proceedings, we summarize the properties of the large rotating disk population at $z \sim 2$ and review the implications of this population for galaxy assembly.  These proceedings are organized as follows:  In \S\ref{sec:prop}, we present the main properties of the large rotating disk population revealed in our SINFONI integral-field spectroscopic data (originally presented in \citealt{For+06_KLS,Gen+06_KLS,Bou+07_KLS,Sha+08_KLS}).  In \S\ref{sec:cold}, we examine the fueling mechanisms for the high star formation rates in these systems and find that the smooth but rapid accretion of cold gas is the only mechanism consistent with the observations (see also \citealt{Gen+06_KLS,Sha+08_KLS}).  In \S\ref{sec:bulge}, we show that detailed dynamical modeling of these galaxies reveals an evolutionary sequence through which young classical bulges are secularly formed (originally presented in \citealt{Gen+08_KLS}).  Finally, in \S\ref{sec:conclu}, we conclude.

\section{Properties of \boldmath $z \sim 2$ Star-Forming Disks}
\label{sec:prop}

\subsection{Characteristics of the Population}
\label{sec:prop_char}

Perhaps the most surprising result of the SINS survey was the discovery of a significant population of large ($r_{1/2} \sim 5-10$ kpc) rotation-dominated objects.  The shape and amplitudes of the rotation curves in these galaxies are consistent with those measured in ionized gas in local spiral galaxies \citep{Bou+07_KLS}.  Fourier decomposition (kinemetry) of the velocity and velocity dispersion maps of these galaxies and comparison with local systems establishes quantitatively that these galaxies have ``spider diagrams" consistent with those observed in local disk galaxies \citep{Sha+08_KLS}.

However, though dynamically similar to local spiral galaxies, high-redshift disks have many characteristics unparalleled in the $z=0$ Universe.  The spatial distribution of H$\alpha$ emission reveals a marked ``clumpiness" of the star formation activity into regions of FWHM size $= 1-3$ kpc \citep[e.g.][]{Gen+08_KLS}.  Broad-band {\it Hubble Space Telescope} imaging of these systems illustrates that these features are also prominent in the stellar distribution (\citealt{ElmElm06_KLS}; F\"orster Schreiber et al.~in preparation).  Dynamical and spectral energy distribution fitting converge on masses of these super-star-forming clumps of $10^8-10^9\ M_\odot$ \citep{Gen+06_KLS,Elm+09_KLS}, a few percent of the total galaxy stellar mass (typically $5 \times 10^{10}\ M_\odot$).  The typical $z \sim 2$ disk galaxy contains $8-10$ such clumps \citep{Gen+06_KLS,Elm+09_KLS}.

Observations of both face-on and edge-on systems suggests that these clumps are roughly spherical, implying a large scale-height in the disks ($h_z \sim 1$ kpc, versus $r_{1/2} \sim 5-10$ kpc; e.g. \citealt{For+06_KLS,ElmElm06_KLS,Gen+08_KLS}).  This is confirmed in the large velocity dispersions observed with SINFONI; detailed dynamical modeling accounting for projection and observational effects indicates the typical random motions are large, $v/\sigma \sim 1-7$ (\citealt{Gen+08_KLS}; Cresci et al.~submitted).  High-redshift disks are therefore much thicker than their low redshift counterparts ($v/\sigma \sim 10-20$), and the implications of this difference are examined in \S\ref{sec:bulge} below.

Finally, high-redshift disks also exhibit much larger star formation rates ($\sim 30-200\ M_\odot\ {\rm yr}^{-1}$) than local disk galaxies, indicative of the high gas fractions in these systems \citep{For+06_KLS,Gen+06_KLS,Gen+08_KLS}.  In the local Universe, such high SFR are nearly always associated with a recent major merger; however, at high redshift, kinemetry analysis has shown these galaxies to have kinematic properties inconsistent with recent major interactions \citep{Sha+08_KLS}.  Moreover, modeling of the spectral energy distributions of these galaxies indicates that the current SFR has been roughly constant over at least $0.5$ Gyr in these systems (\citealt{Gen+06_KLS,Dad+07_KLS}; F\"orster Schreiber et al.~in preparation).  Such long-lasting high SFR and gas fraction suggests that the fueling of high SFR at high redshift occurs very differently than that at low redshift (see discussion in \S\ref{sec:cold}).

\subsection{The Appearance of Local Scaling Relations}

The above evidence suggests that high-redshift disks, given their clumpiness and thickness, cannot passively evolve into their low-redshift late-type disk counterparts (see \S\ref{sec:bulge}).  Nevertheless, their rotation curves obey the same radius-velocity relation as observed at $z=0$ \citep{Bou+07_KLS}.  It is therefore of interest to examine scaling relations at $z \sim 2$, in order to probe the fundamental physics governing these relations.

Despite $z \sim 2$ systems having similar rotation velocities to local disk galaxies, detailed dynamical modeling of these systems coupled with spectral energy distribution analysis reveals that high-redshift disks are significantly offset from the local Tully-Fisher ($M_*-v$) relation (Cresci et al.~submitted).  This may reflect the high gas fractions in these $z \sim 2$ disks relative to local disks, or it may reflect a fundamental evolution of this relation with time.

In contrast, the high gas fractions and gas surface densities present at $z \sim 2$ drive star formation with an efficiency consistent with a universal Schmidt-Kennicutt scaling relation at high and low redshifts \citep{Bou+07_KLS}.  The presence of super-star-forming clumps in $z \sim 2$ disks thus does not seem to fundamentally alter the physics of star formation within these galaxies.

In cosmological models, this rapid star formation is often associated with rapid accretion onto a growing supermassive black hole (SMBH), a process that should be apparent via broad emission lines in our SINS galaxies.  Indeed, stacking of our SINFONI spectra for all objects reveals a broad component underneath the H$\alpha$-[N{\small II}] complex (Shapiro et al.~in preparation).  If this feature is interpreted as evidence of active galactic nuclei in the SINS galaxies, SMBH masses can be inferred from the luminosity and FWHM of the broad line.  These SMBH masses are offset from local relations by an order of magnitude, in that the SMBH are under-massive for their host galaxies, implying a delayed assembly of black holes in $z \sim 2$ disk galaxies (Shapiro et al.~in preparation).

It seems, then, that some processes (star formation and the physics governing the radius-velocity relation of galaxy disks) are independent of redshift, while those involving the gradual growth of stellar and SMBH mass are not preserved out to $z \sim 2$.

\section{Creating Clumpy Disks: Cold Flows vs. Major Mergers}
\label{sec:cold}

\subsection{Differences from Major Merger Remnants}

Given the lack of low-redshift analog for the clumpy, thick, rapidly star-forming disks seen at $z \sim 2$, the formation of these high-redshift objects begs explanation.  The traditional paradigm through which systems with high SFR were explained was via gas-rich major mergers.  Indeed, it has been known for some time that a merger with a sufficiently large gas fraction can result in very high SFR and in a gas-rich remnant disk \citep[e.g.][]{BarHer96_KLS}.

However, this almost certainly is not the explanation for the high-redshift disks seen in our SINS sample.  As described above, the SINS high-redshift disks show no kinematic evidence of recent disturbances and have likely experienced constant SFR over at least $0.5$ Gyr, unlike the bursty star formation histories associated with galaxy interactions.

Several other key features in the data argue against a major merger origin for these systems.  Most critically is the lack of central mass concentrations in some systems.  While a major merger always creates a central concentration of mass, a significant fraction of high-redshift disks exhibit mass concentrations that peak in rings located $2-8$ kpc away from the galaxy center \citep{Gen+08_KLS}, a configuration that cannot readily be created in a major merger.

Another relevant characteristic of the high-redshift disks is the observed clumpiness; the $\sim 10$ clumps in a galaxy, each with a few percent of the galaxy's mass, together make up a dynamically important component in these systems.  With this significant fragmentation, high-redshift disks do not resemble the smooth and idealized remnants of simulated gas-rich major mergers; it is not obvious that such a merger could simultaneously produce gas-rich and fragmented disks, as are observed.  In contrast, these mergers probably correspond to the sub-mm population detected at $z \sim 2$, whose compact and bursty star formation history almost certainly results from a major merger of two gas-rich systems \citep{Tac+08_KLS}.

\subsection{Evidence for a Smooth Accretion Mechanism}

Consequently, an alternative mechanism must fuel the constant high SFR seen in high-redshift disks.  A natural explanation can be found in the ``cold" (with respect to the virial temperature) gas flows entering these (high-$\sigma$ peak) haloes along filaments.  Such accretion has been shown to penetrate through to the center of the halo, replenishing the gas reservoir within the galaxy in a smooth (average merger ratio $>$ $10$:$1$) manner \citep[e.g.][]{DekBir06_KLS}.  This picture is in keeping with the SINS observations of kinematically undisturbed disks forming stars at constant and very high rates \citep{Gen+06_KLS,Sha+08_KLS}.

Statistical studies of high-redshift populations are also converging on this picture with independent arguments.  The detection of a tight SFR-$M_*$ relation from $z \sim 0-2$ implies that the primary driver for star formation cannot be a bursty mechanism such as a series of major mergers and instead must be a smooth mass-dependent process \citep{Noe+07_KLS,Dad+07_KLS}.  Additionally, analysis of the accretion histories of dark matter haloes in the Millennium simulation indicates that only a small fraction of galaxies with the masses and star formation rates of the SINS high-redshift disks would be expected to have undergone a recent major ($<$ $3$:$1$) merger \citep{Genel+08_KLS}.

\section{Evolution of Clumpy Disks and the Formation of Bulges}
\label{sec:bulge}

In addition to revealing the importance of the cold flow accretion mechanism in galaxy formation at high redshift, the SINS galaxies have also provided detailed insight into the modes of galaxy and bulge growth in the high gas fraction, high turbulence regime.  In particular, the observed clumpiness can be understood simply as the characteristic Jeans length in these systems \citep{Gen+08_KLS}.  The expectation for a gas-rich disk to fragment in this manner has also been borne out in hydrodynamical simulations of this process, which generate systems remarkably similar to the observations \citep[e.g.][]{Imm+04_KLS,BouElmElm07_KLS}.

Once present, the dynamical timescale of these clumps in such highly turbulent, gas-rich disks implies that they should migrate to the center of the disk within $\sim 1$ Gyr \citep{Gen+08_KLS}.  The coalescence of several such clumps would then be expected to produce a young classical bulge, despite its secular origins \citep{ElmBouElm08_KLS}.  Indeed, such features are observed in some of the SINS galaxies (\citealt{Gen+08_KLS}; F\"orster Schreiber et al.~in preparation).  Moreover, the mass of the central concentration correlates well with the metallicity and therefore chemical age of the galaxy; older galaxies in which clumps would have had more time to migrate and merge are observed to contain the expected massive bulges \citep{Gen+08_KLS}.

These bulges can then stabilize the remaining disk against the formation of future super-star-forming clumps \citep{DekSarCev09_KLS}, such that the remaining disk, supplemented by subsequent gas accretion, exhibits a smooth exponential surface brightness profile \citep{BouElmElm07_KLS}.  In this manner, the high-redshift disks can evolve into local bulge-dominated systems (e.g. elliptical, lenticular, and Sa galaxies).  This scenario is in keeping with the similar number densities of the high-redshift disks and their probable descendants and with predictions based on cosmological simulations \citep[e.g.][]{Con+08_KLS,Genel+08_KLS}.

\section{Conclusions}
\label{sec:conclu}

From the SINS survey, and from complementary observational and theoretical campaigns, a new understanding of the formation and evolution of massive galaxies has emerged.  In particular, a substantial fraction of these systems are now known to undergo a disk-like stage, in which the galaxy consists of a large, rotating, thick, clumpy disk.  Moreover, these disks are observed to form stars rapidly and to do so at a continuous rate for up to and exceeding one Gyr.

In contrast to local objects with high SFR, these galaxies are inconsistent with a major merger origin; evidence against a recent major merger is apparent in both their star formation histories and their internal dynamics.  Instead, the high SFR in $z \sim 2$ star-forming disks is likely driven by the smooth but rapid inflow of gas along filaments in the cosmic web.

The gas-rich disks that result from this process are globally unstable and are observed to collapse into kpc-scale clumps, in keeping with theoretical expectations.  With time, these clumps migrate to the center of the potential and combine to form a nascent bulge, a process that is directly observed in the SINS data.  At $z=0$, the eventual product of this system will be a bulge-dominated galaxy, whose central mass concentration was generated without a major merger.

\acknowledgements
The SINS project would not have been possible without the helpful and enthusiastic support of the ESO staff, particularly at Paranal Observatory, over the many observing runs and several years during which these observations were carried out.  We also thank the SINFONI and PARSEC teams for their hard work on the instrument and the laser, which allowed our program to be so successful.


\end{document}